\documentstyle[12pt,aasms4,psfig]{article}

\def\ltw{\>\hbox{\lower.25em\hbox{$\buildrel <\over\sim$}}\>}
\def\gtw{\>\hbox{\lower.25em\hbox{$\buildrel >\over\sim$}}\>}

\def \bold#1{{\bf #1}}
\def \rhat{\hat \bold r}
\def \xhat{\hat \bold x}
\def \yhat{\hat \bold y}
\def \zhat{\hat \bold z}
\def \csalph{\cos \alpha}
\def \snalph{\sin \alpha}
\def \cjunk{\cos [(r-r_*)/r_L]}
\def \sjunk{\sin [(r-r_*)/r_L]}
\begin{document}

\vglue 1.0 in

\title{\bf THE SHAPE OF PULSAR POLAR CAPS}

\vskip 0.3in
\author{\bf P. N. ARENDT, Jr.}

\centerline{ and}

\vskip 0.2in
\author{\bf J. A. EILEK}

\vskip 0.1in
\affil{Physics Department and Astrophysics Research Center \\
New Mexico Tech, Socorro NM 87801, USA}

\begin{abstract}

Rotation distorts the vacuum magnetic field of a pulsar from that of a
simple dipole.  The effect is particularly strong close to the light
cylinder, but also affects the field close to the stellar surface.  We
find the shape and locus of the field lines which just close at the light
cylinder.  Their footpoints define the pulsar polar cap.  We find this
cap is asymmetric and distorted.  We also find that the polar cap is
{\it not} strongly elliptical.  This result disagrees with
calculations based on  the non-rotating dipole field.  We present our
numerical results, and discuss consequences for interpretation of mean
profiles and for pulsar statistics.

\keywords{pulsars, magnetic fields}

\end{abstract}

\section{Introduction}

The pulsar polar cap is the region mapped out on the stellar surface
by the footpoints of magnetic field lines which do not close within
the light cylinder (open field lines).  It is bounded by the ``last
closed field lines'', those lines which just turn around  at the light
cylinder and begin there to return to the star. 
Plasma flowing out from the polar cap is very likely to be the source
of coherent, pulsed radio emission.  Any effort at
interpreting pulsar radio profiles must therefore be founded on an
understanding the nature of this polar cap. 

It  was recognized early on that basic pulsar observations could be
simply explained by a dipolar magnetic field centered on the star.
Such a field produces a relatively small open field line region, which
is consistent with the duty cycle of mean profiles.  In addition, this
magnetic geometry can nicely explain the observed rotation of
polarization position angle across the pulse. 
Sturrock (1971) pointed out that the angular size of this open field
line region, in an aligned dipole, is given by
\begin{equation}\label{eq:rhoPC}
\rho_{\rm PC} \simeq \left( { R /  r_{\rm LC}} \right)^{1/2}
\end{equation}
where $r_{\rm LC} = c / \Omega$ is the light cylinder radius, and
$\Omega$ is 
the angular rotation frequency.  Does this extend to a tilted dipole?
There were some early suggestions ({\it e.g.} Narayan \& Vivekanand
1982) that the radio emission region is elongated in the meridional 
direction, but this idea was not supported with larger data sets ({\it e.g.}
Lyne \& Manchester 1988).  In fact, most subsequent  interpretation of
pulsar radio profiles has assumed a polar cap which is circular, with
a characteristic size given by equation (1) ({\it
e.g.}, Backer 1976; Lyne \& Manchester 1988; Rankin 1990 and
references therein; Gil, Kijak \& Seiradakis 1993 and references
therein).  A circular polar cap seems to be consistent with some
striking trends found in the data, particularly the period-pulse width
relations established by Rankin for coral and conal pulsars.  

The shape of the polar cap {\it on the star's surface} turns out to be
very sensitive to the shape of the magnetic field lines {\it close to
the light cylinder}.   If one assumes the magnetic field is that of a
pure, static dipole, aligned at an angle $\alpha$ to the rotation
axis, the resulting polar cap is not circular. Roberts \& 
Sturrock (1972), and also Biggs (1990), found that 
the polar cap in an oblique pure dipole is more elliptical than circular.
It has a longitudinal size given by equation (\ref{eq:rhoPC}), but
is compressed in the  
latitudinal direction, by an amount up to $\sim 60$\% in an orthogonal
rotator.   

Rotation and relativity change all this.  A rotating dipole is in fact
a time-dependent source for the magnetic field.  Basic physics leads
us to expect that close to, and beyond, the magnetic field must change
from the static field, which is the near-field solution, to the
far-field, radiative solution  (in the terminology of Jackson 1975). 
Deutsch (1955) solved this problem, and showed that the field of a  
misaligned rotating dipole does, indeed, have field lines which are strongly
distorted close to the light cylinder.  Romani \& Yadigaroglu (1995) noted
that these high-altitude distortions will connect back to the stellar
surface, resulting in a polar cap that is neither circular nor simply
elongated.  

Our contribution in this paper is to display the effect of
relativistic field line distortion on the shape of the polar cap at
the stellar surface, and in the radio emission region. 
We locate and map
the last closed field lines numerically, and from this we find the exact
shape of the polar cap for a rotating ``pure'' dipole.  We find that the
polar cap is much closer to circular than was previously thought, and
that it is also asymmetric, with interesting distortions on the
leading edge.

\section{Magnetic Field of a Rotating Dipole}

Deutsch (1955) found that relativity distorts the magnetic field lines
of a rotating, inclined dipole, compared to those of a pure dipole.
Even small inclination angles (let $\alpha$ be the angle between the
rotation and magnetic 
axes) add a toroidal component to the purely poloidal dipole
field. The distortions become quite dramatic for large inclination angles. 
 Examples of the distortions can be seen in Romani \&
Yadigaroglu (1996), and  in Higgins \& Henriksen (1997), as well
as the figures in this paper.

In order to study the magnetic field structure, we reproduce the
Deutsch solution using a simpler analytic approach.  We mix a static
dipole, parallel to the rotation axis, and two oscillating ones,
perpendicular to the rotation axis.  (We have put the details in the 
Appendix.)   The region well inside the light cylinder corresponds to
the ``near zone'', or static, solutions in Jackson's (1975) terminology,
and the region well outside is the ``far zone'', or radiation, solutions.
The region around the light cylinder is the ``intermediate zone'',
and we expect interesting field structure there.

We developed a line-tracing code to explore the solutions.  We
start at the star's surface, picking points with polar angle (relative
to the magnetic axis) equal to $\rho_{\rm LC}$, the
standard polar cap radius, and uniformly spaced in azimuth around the
magnetic axis.  We use Runge-Kutta integration to follow a
particular field line in space.  We determine whether that line closes
inside the light cylinder or crosses it.  We then iterate on the
starting polar angle (which fixes the footpoint of the line) until we
have found the field line which just closes at the light cylinder.  We
call this the last closed field line.  

In Figures 1 through 4 we show plots of the last closed field lines for
four inclination angles.  All simulations assume a one second rotation
period; the light cylinder is at 4800 stellar radii for a typical
stellar radius of 10 km.
. In each figure,
we show two, two-dimensional projections, and a three-dimensional
view. Distances are scaled to the light cylinder radius.   
The first two-dimensional 
view is projected onto the ``rotation equator'', {\it i.e.} looking
down the rotation axis.  Rotation is in the counter-clockwise
direction.  The second two-dimensional view is projected onto the
``magnetic equator'', {\it i.e.} looking down the magnetic axis.  In
each three-dimensional view, the 
rotation axis is indicated by a vertical line.  The distortions due
to rotation are apparent.   Even at small angles $\alpha$, the field
lines are twisted out of the meridional planes (compared to a pure
dipole field).  At larger angles, field lines are ``swept back''; lines
which start equally spaced around the magnetic axis become packed
together at higher altitudes.  

We note that this high altitude distortion and sweep-back of field
lines arise in a purely vacuum solution.  Thus, these effects do not
depend on arguments about plasma inertia close to the light cylinder.
Rather, they are simple consequences of relativity. 

\section{ Consequences for the Polar Cap}

We also use this code to explore the shape of the polar cap.  In
Figures 5 and 6 we show the footpoints of the last closed field lines,
for a range of inclination angles.  In each figure, we show the
classical polar cap in a light dotted line, from equation (1), and the true
polar cap, in a dark line.  Rotation is to the right in each figure. We
find two striking results.   

First, the polar cap  is not noticeably elongated.  It does not
deviate much from a circular shape at the classical radius $\rho_{\rm
PC}$ (indicated by dotted lines in the figures), and the deviations
that exist are asymmetrical in azimuth.  This result disagrees with
the elongated polar cap obtained with an oblique, pure dipole.  We
have not come up with any simple explanation.    We can only note 
that field lines which start on the top and bottom (north and south)
of the polar region extend further away from the rotation axis than
they do in the nonrotating case.

Second, the polar cap is asymmetrical, both north-south and east-west.
A striking discontinuity develops on the north leading side for
inclination angles $\sim 40^{\circ} - 60^{\circ}$, being strongest for
angles $\simeq 45^{\circ}$.  This discontinuity
is a consequence of the rotation-induced sweepback of field lines,
which causes the last closed field
lines to diverge at high altitudes.  This effect can be seen
in Figures 1 through 4.

We have investigated the nature and origin of this asymmetric shape,
particularly the ``glitch'' at intermediate inclination angles.
Our representation of the 
field (as given in the Appendix) allows us to separate out the pure
dipole part, $\bold B_{\rm dip}$, from the total field.  At small
altitudes, the deviation $\Delta \bold B = \bold B - \bold B_{\rm
dip}$ is small.  It is easy to show that  $\Delta B / B \sim O ( r /
r_L)$ for $r \ll r_L$.  Thus, both the
magnitude and the direction of the field vary little at low
altitudes compared
to the field of a pure dipole.  In particular, there is no
low-altitude anomaly at the magnetic azimuth of the glitch, despite
its discontinuous appearance.

We also find that the polar cap shape does not depend on altitude, for
altitudes well within the light cylinder.  It is
the same whether on the star's surface or at the altitude of typical
radio emission ($\sim 10$ stellar radii, typically).  It may thus be more
correct, formally, to refer to this shape as a ``radial cross section
of the polar flux tube'' (to distinguish a high-altitude cross section
from the physical open field line footpoints on the stellar surface).
However, in the interest of conciseness, we shall continue to use the
term ``polar cap''.

\section{Discussion and Conclusions}

In this paper we have revisited the question of the shape of the polar
cap in a rotating dipole.  We find that the polar cap is close to
circular, and its scale is well approximated by the  classical radius,
$\rho_{\rm PC}$.  Contrary to previous work which ignored the relativistic
effects of rotation, we do not find an elongated polar cap. In
addition, the polar cap does show interesting asymmetries,
especially the ``glitch'' on its northeast (leading, pole-ward) side.  

Will this structure have direct consequences for observations of
individual pulsars?  We suspect not.

The glitch is due to the high-altitude structure of the field lines.
Close to the surface, the field does not deviate significantly from a
pure dipole, even in the region of the glitch.  Thus, we expect little
signature of this structure in 
polarization position angle sweeps. One might anticipate
an irregular plasma flow in the region of the glitch, reflecting the
higher-altitude field line divergence.  Flow irregularities may give
rise to irregularities in the radio emission.\footnote{This statement
is no more than speculation, however, until backed up by a good radio
emission model.} Lines of sight which pass close to this glitch -- say
at an impact 
parameter 1/3 to 1/2 of $\rho_{\rm PC}$, might be expected to result in
strongly asymmetric mean profiles (with either leading or lagging
emission enhanced).   Separating this out from other causes of profile
asymmetry seems a difficult game, however.  A glance at any modern data set
-- for instance the 21 cm profiles taken by Rankin, Stinebring \&
Weisberg (1989), or the multifrequency profiles presented by Eilek,
Hankins \& Rankin (1998) -- shows that symmetric profiles are rare.
In terms of the magnetic geometry of the polar cap, nearly all radio
emission seems to be  asymmetric in magnetic azimuth.  

Another caveat in connecting these models to observations is the
question of the filling factor for radio emission.  Most work on
interpreting observations has implicitly assumed that the polar cap is
uniformly filled, at least in magnetic azimuth.  Theoretical models do
not necessarily predict this, however,  For instance, the steady-state
models of Arons \& Scharlemann (1979; also Scharlemann, Arons \&
Fawley 1978) find that pair formation should occur only in the ``top
half'' of the polar cap region (the half toward the magnetic axis).  If we
associate pair formation with  radio emission, this model predicts an
elongated emission region, smaller than the classical $\rho_{\rm PC}$
size.  

Thus, while a serious discussion of observational
consequences must await understanding of the emission physics in this
region, we doubt that dramatic consequences will result for individual
pulsars.  We do expect, however, that our results will impact
discussions of pulsar statistics.  In particular, various authors have
discussed polar caps which are elongated in either the north-south
direction (as suggested by Narayan \& Vivekanand 1982), or the
east-west direction (as predicted by an oblique, simple dipole). These
elongations can in principle be detected by looking at the
distribution of pulse widths, $W$, against polarization sweep rates.
The second of these depends on the impact angle ($\delta$, the angle
between the line of sight and magnetic axis), and the first depends on
the polar cap width at the particular impact angle.  Recently, Biggs (1990)
and McKinnon (1993) have argued that the statistics of $W$, $\delta$
and the inferred inclination $\alpha$ are more consistent with a
flattened polar cap than a circular one.  Their conclusions are not
consistent with our result. The discrepancies may be telling us that
polar caps are not uniformly filled with radio emission.
Alternatively, they may be telling us that current samples of
well-observed pulsars may not be unbiased in impact and inclination
angles.

\section*{}
We enjoyed discussions during the course of this work with Don
Melrose, Jim Weatherall, Tim Hankins and  David Moffett. 
This work was partly supported by the NSF through grant AST-9315285. 

\appendix
\section{The Magnetic Field of a Rotating Dipole}

We present a new derivation of the Deutsch (1955) solution for the
magnetic field of a rotating dipole.  We wish to produce the field due
to a dipole, with amplitude $\bold m$, which is inclined at an angle
$\alpha$ to the rotation axis.  We 
represent this as a sum of a static dipole, in the $\hat \bold 
z$ direction, and two out-of-phase oscillating dipoles, in the $\hat
\bold x$ and $\hat \bold y$ directions. That is, we superpose the
field due to an static dipole, with amplitude $m \cos \alpha$, in the
$z$ direction with that due to an oscillating dipole in the 
$x$ direction, with amplitude $m \sin \alpha$,  and another
oscillating dipole, also with amplitude $m \sin \alpha$ in the $y$
direction, 90 degrees out of phase with the $x$ dipole.  

The field of a static dipole is, of course, (Jackson, 1975, eq. 5.56)
\begin{equation}
{\bold  B_{\rm dip}(r)} = {\bold 3 \rhat (\rhat \cdot \bold m )
- \bold m \over r ^3} 
\end{equation}
The field for a time-varying magnetic dipole is
given by (Jackson 9.35)
\begin{equation}
{\bf B_{\rm osc}(r)} = k^2 {\bf \left(\rhat \times m \right)\times \rhat}
{e^{ikr}\over r} +  \left[3{\bf \rhat (\rhat \cdot m) - m} \right]
\left( {1 \over r^3} - {i k \over r^2} \right) e^{ikr} 
\end{equation}
where ${\bf B}$ and ${\bf m}$ are understood to carry a time-dependence
$e^{-i \Omega t}$, and  the real part is to be taken.  Here, $\Omega$
is the angular frequency of rotation, so $k = \Omega / c$ is just
$1/r_L$.

 We wish to produce
a dipole which rotates counterclockwise as viewed from ``above''.  As
the real part of $e^{-(i \Omega t)}$ is $\cos (\Omega t)$, we add the
$y$ oriented field with an added factor of $i$, giving it $\sin (
\Omega t)$ dependence.  However, we are really only interested in the
field for a fixed time, so we replace all phase factors by the factors
$\cos (r-r_*)/r_L$ and $\sin (r-r_*)/r_L$.  This fixes the magnetic pole
at the stellar surface to be in the direction $(\sin \alpha, 0, \cos
\alpha)$, and accounts for the relatively minor time delay from the
origin to the stellar surface.

After superposition of all three fields, and taking the real part of all
terms, we arrive finally at the result.  In 
Cartesian components aligned with the rotation axis, it is
\begin{eqnarray}
\bold B (\bold r) & = &   m \csalph  \bold F_1~ {1 \over r^5} 
~+~  m \snalph ~ \bold F_2 ~{\cjunk \over r^3 r_L^2} 
\nonumber
\\
& & +~  m \snalph ~ \bold F_3 {\sjunk \over r^3 r_L^2}  
\nonumber
\\
& &+ ~ m \snalph  ~ \bold F_4 \left[ {\cjunk \over r^5} - {\sjunk \over
r^4 r_L} \right] 
\nonumber
\\
&& + ~  m \snalph ~ \bold F_5 \left[ {\sjunk \over r^5} + {\cjunk \over r^4
r_L} \right]
\end{eqnarray}
with
\begin{eqnarray}
\bold F_1 &= &\left[3 x z \xhat + 
3 y z \yhat
+  (3 z^2 - r^2) \zhat \right] 
\nonumber\\
\bold F_2 &= &\left[(r^2 - x^2) \xhat -xy \yhat -  xz
\zhat \right] 
\nonumber\\
\bold F_3 &= &\left[-xy \xhat + (r^2 - y^2) \yhat
 -yz \zhat \right] 
\nonumber\\
\bold F_4 &= &\left[(3 x^2 - r^2) \xhat +  3 x y \yhat +  3 x z \zhat \right] 
\nonumber\\
\bold F_5 &= &\left[3 x z \xhat + 3 y z \yhat \right]
\end{eqnarray}
The first term is that due to the static component, and the rest to
purely rotating components.  The pure dipole field is reproduced by
\begin{equation}
\bold B_{\rm dip} = { m \over r^5} \left[ \bold F_1 \cos \alpha +
\bold F_4 \sin \alpha \right]
\end{equation}
and the extra component added by the rotation is
\begin{equation}
\Delta \bold B = \bold B - \bold B_{\rm dip}.
\end{equation} 

Finally, we note that it is straightforward, but rather tedious, to
verify that this reproduces the Deutsch (1955) solution.  Deutsch
expressed his solution in terms of spherical Bessel functions of the
third kind, $h_1(r/r_L)$ and $h_2(r/r_L)$.  The connection to our form
can be made by noting the expansion of these functions in terms of
$\sin(r/r_L)$ and $\cos(r/r_L)$ ({\it e.g.} Abramowitz \& Stegun 1964).

\clearpage

%  30 degrees double figure
\begin{figure}
\vspace{-0.5in}
\centerline{
\psfig{figure=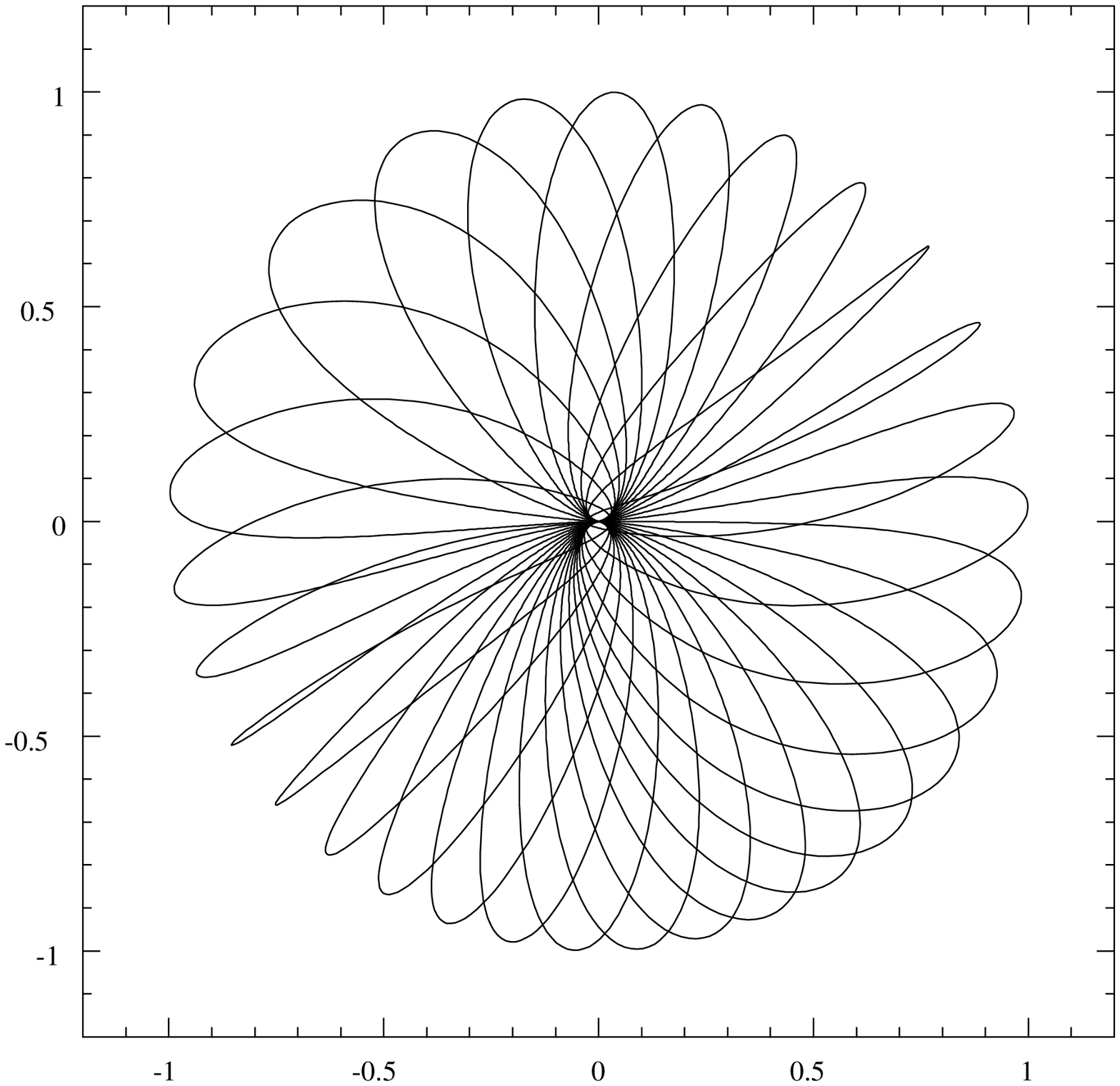,height=7.5cm,width=7.5cm}
\hspace{0.4in}
\psfig{figure=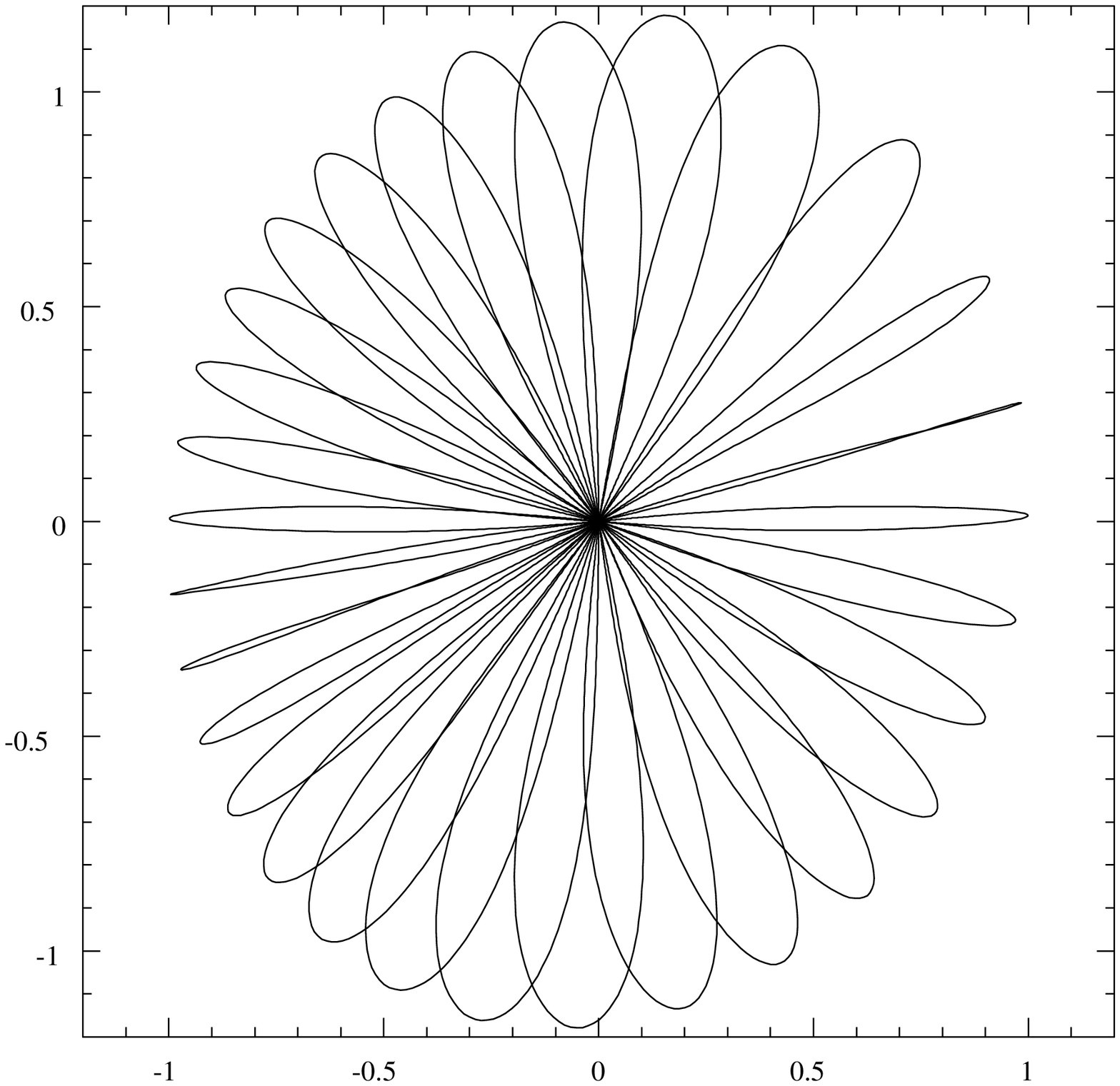,height=7.5cm,width=7.5cm}
}
\vspace{-0.12in}
\end{figure}
\vspace{0.5in}
\begin{figure}
\vspace{-.8in}
\centerline{
\psfig{figure=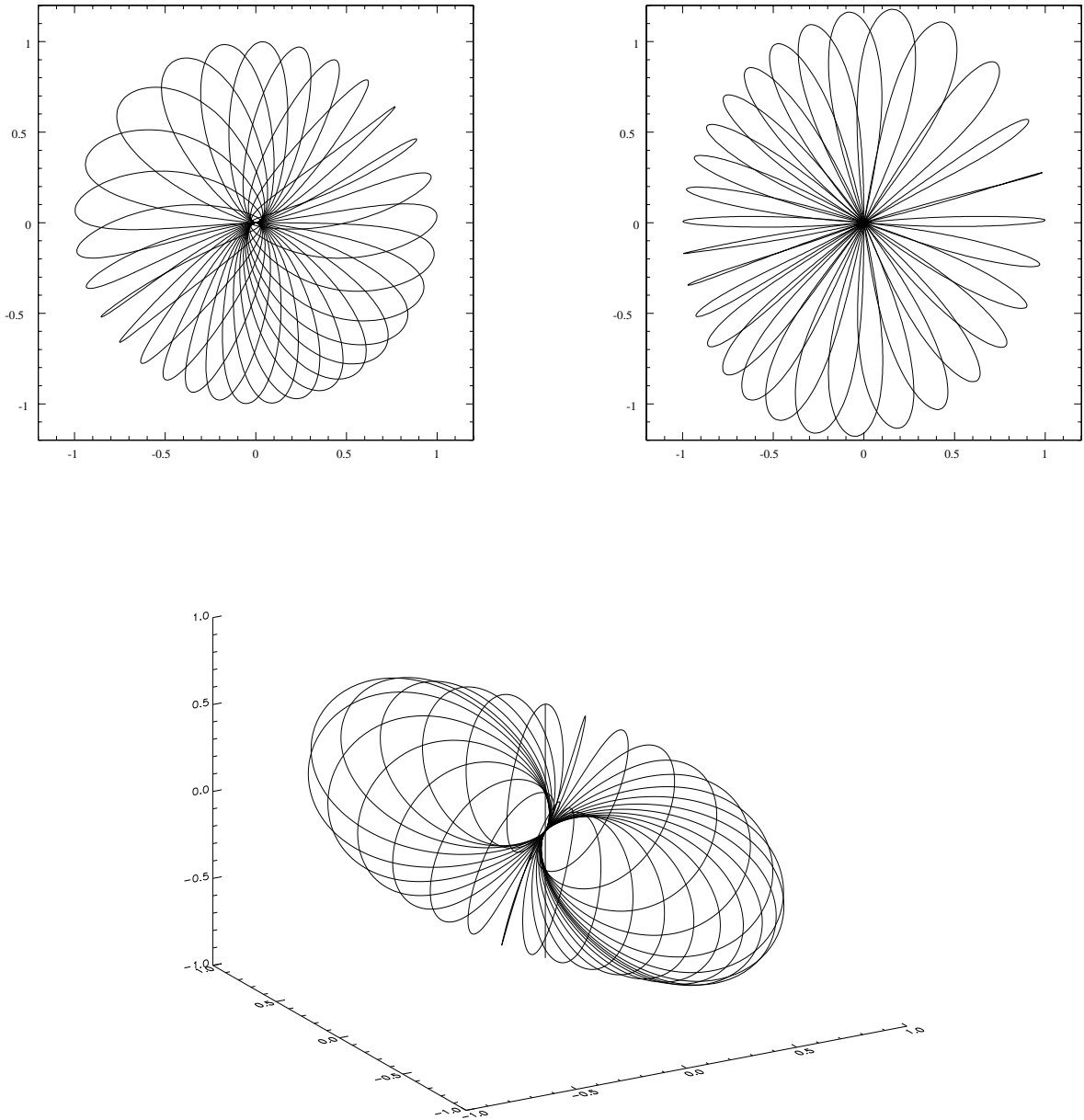,height=9cm,width=12cm}
}
\vspace{-0.12in}
\caption{The last closed field lines for a dipole inclined at
$\alpha = 30^{\circ}$ to the rotation axis.   (a) A projection looking
down the rotation axis; rotation is counterclockwise.  (b)  A
projection looking down the magnetic axis.  (c)  a three-dimensional
view, with the rotation axis indicated by the vertical line. 
%Even at this small inclination, the distortion
%of the field lines from those of a pure dipole is apparent; a toroidal
%component has been added, which causes the field lines to deviate from
%the meridional plane.  
The stellar radius is too small to be shown in
these figures.}
\end{figure}

\newpage
%  45 degrees double figure
\begin{figure}
\vspace{-0.2in}
\centerline{
\psfig{figure=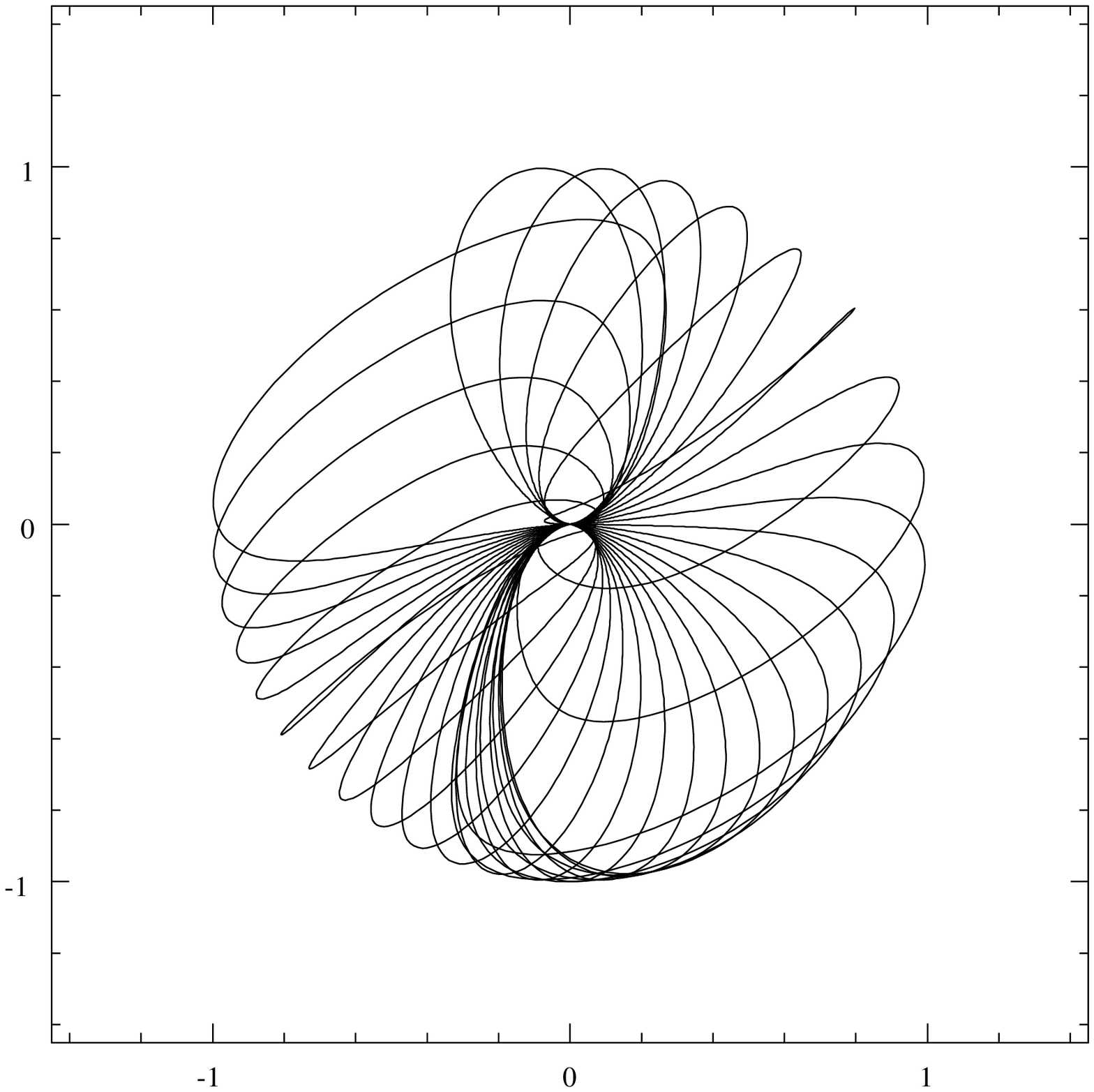,height=7.5cm,width=7.5cm}
\hspace{0.4in}
\psfig{figure=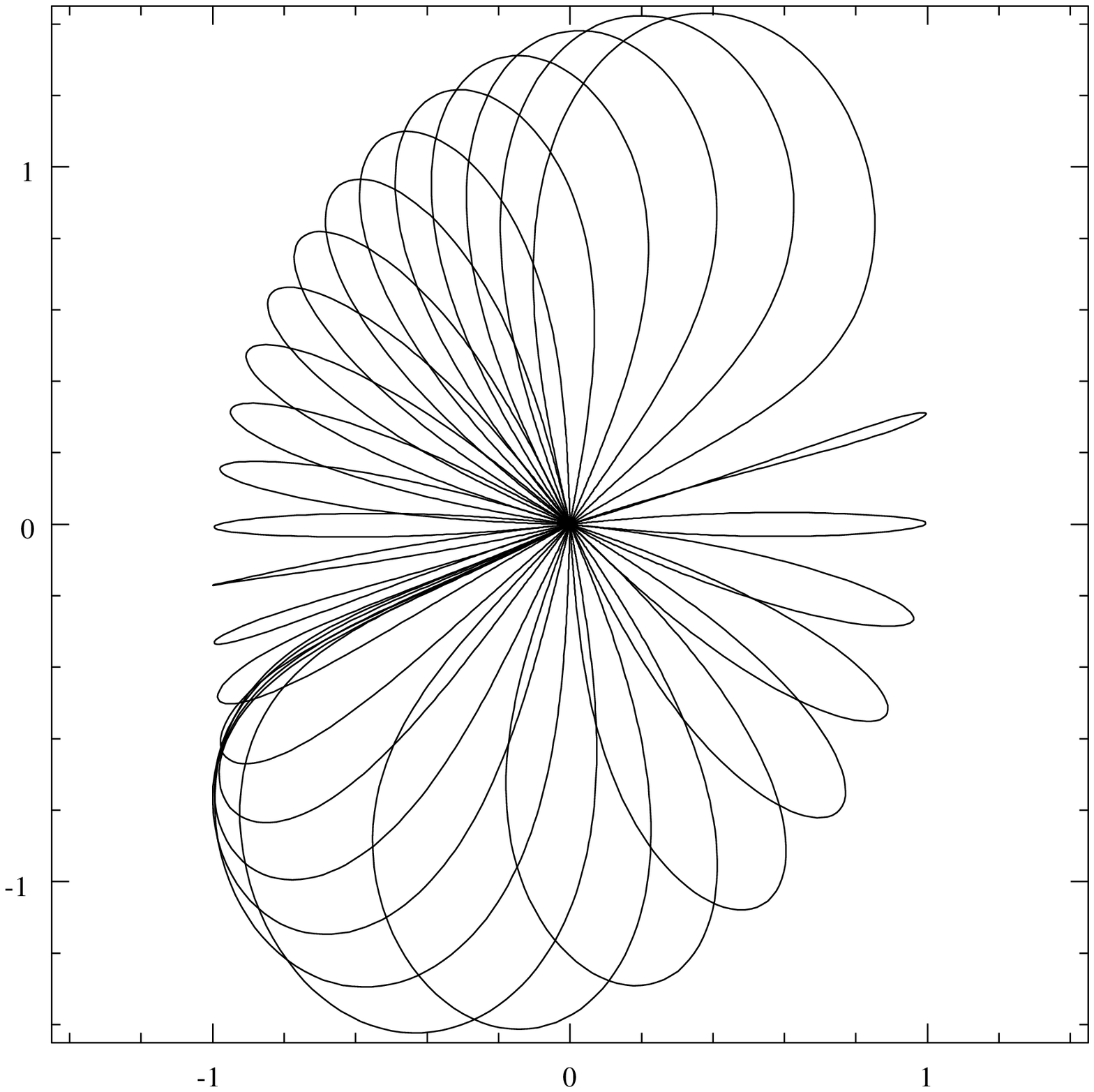,height=7.5cm,width=7.5cm}
}
\vspace{-0.12in}
\end{figure}
\vspace{0.5in}
\begin{figure}
\vspace{-.8in}
\centerline{
\psfig{figure=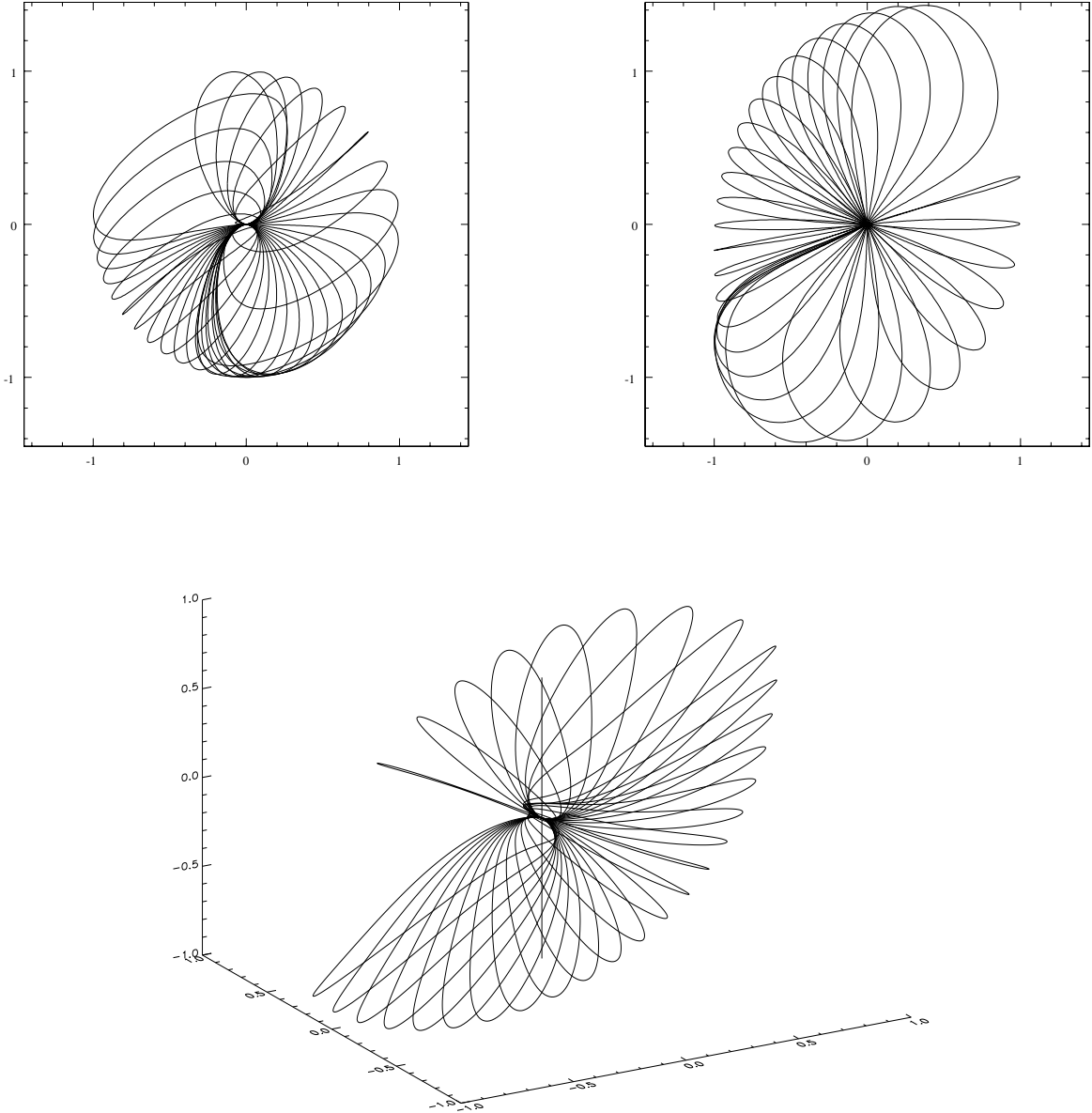,height=9cm,width=12cm}
}
\vspace{-0.12in}
\caption{The last closed field lines for a dipole inclined at
$\alpha = 45^{\circ}$ to the rotation axis.  The layout is the same as
in Figure 1.  The ``sweep-back'' of the field lines, and their
avoidance of one quadrant of the magnetosphere, can be seen here (all
figures in this paper depict field lines which are equally spaced in
magnetic longitude at the stellar surface).}
\end{figure}

\newpage
%  60 degrees double figure

\begin{figure}
\vspace{-0.3in}
\centerline{
\psfig{figure=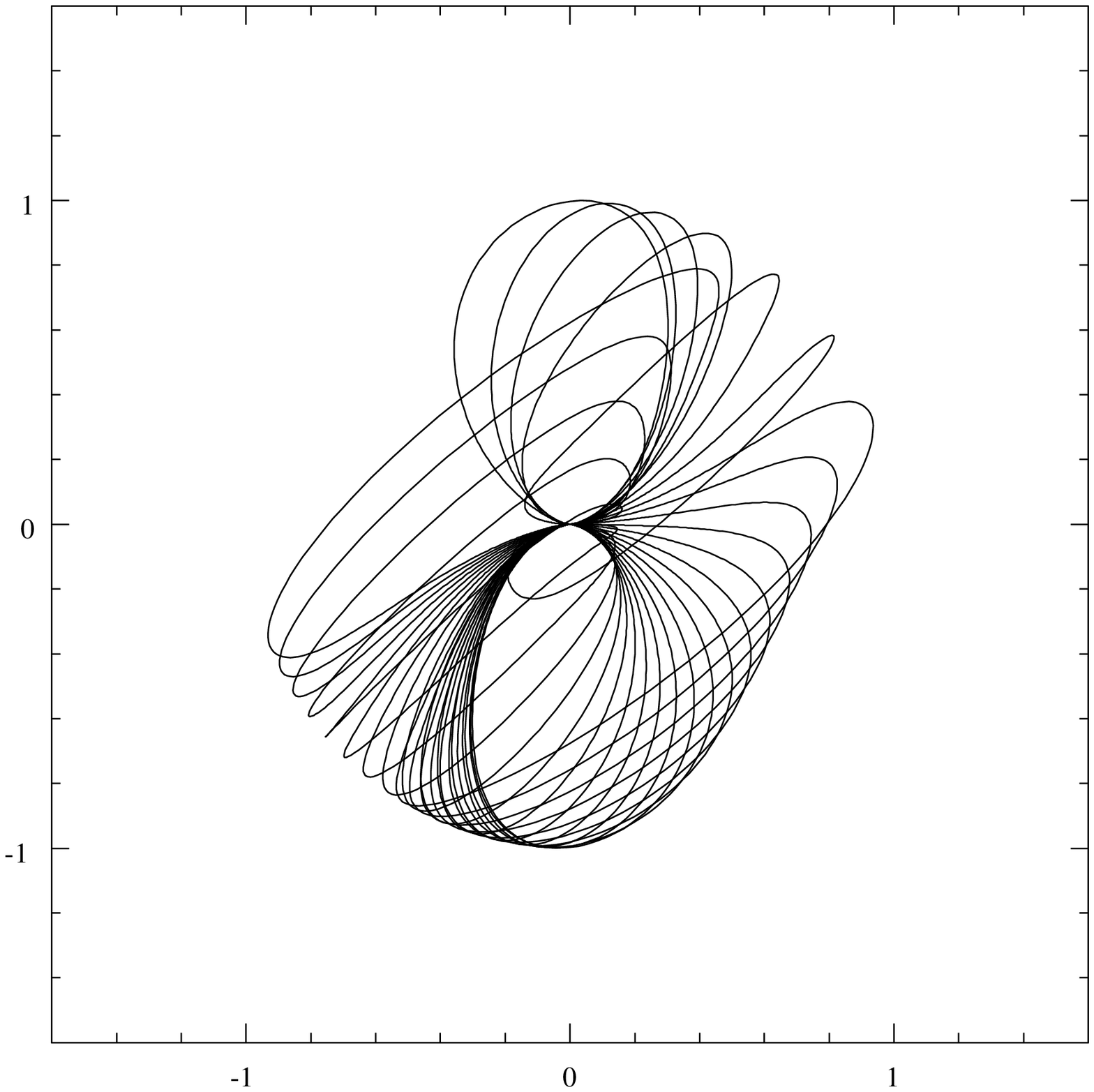,height=7.5cm,width=7.5cm}
\hspace{0.4in}
\psfig{figure=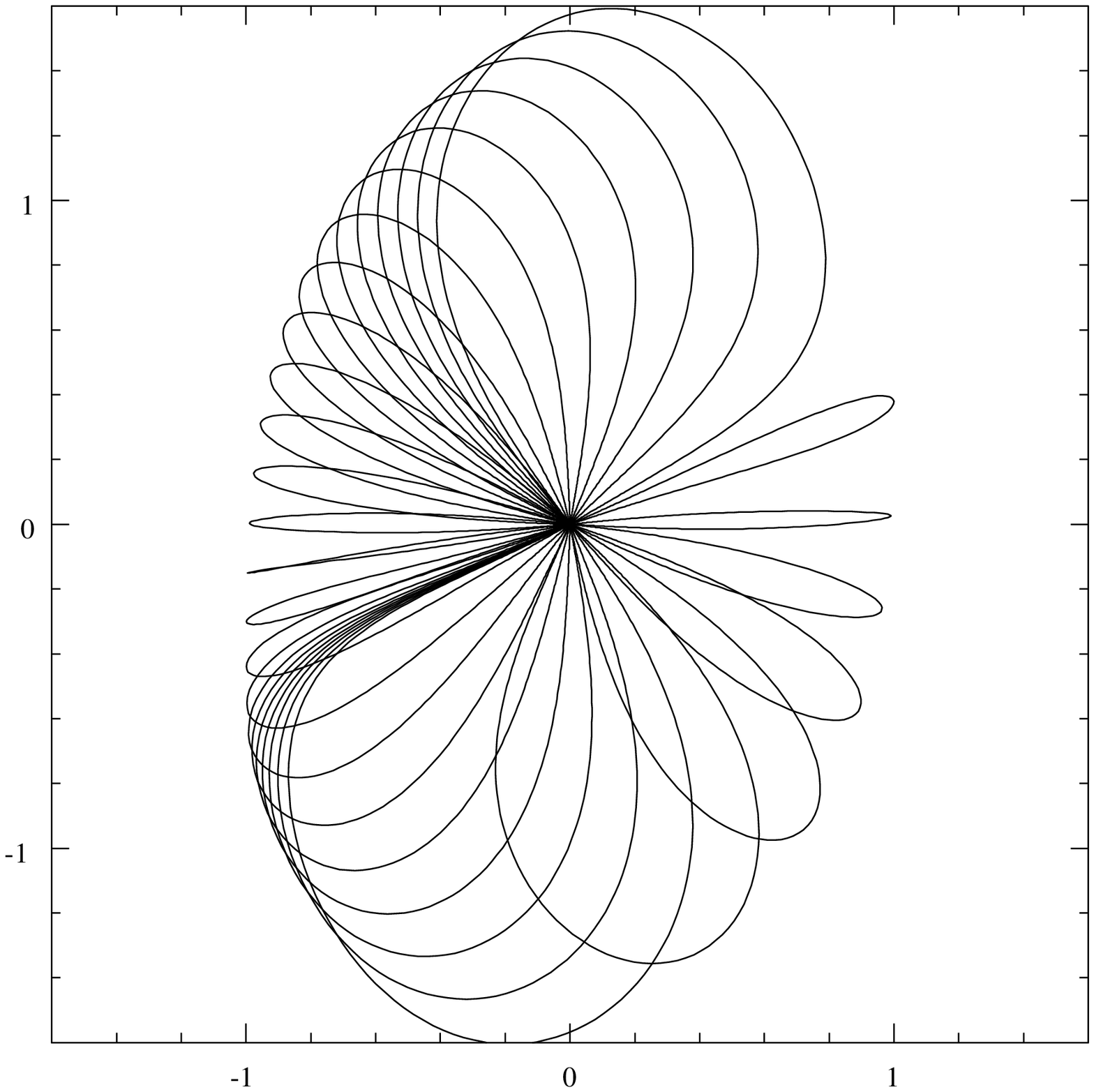,height=7.5cm,width=7.5cm}
}
\vspace{-0.12in}
\end{figure}
\vspace{0.5in}
\begin{figure}
\vspace{-.8in}
\centerline{
\psfig{figure=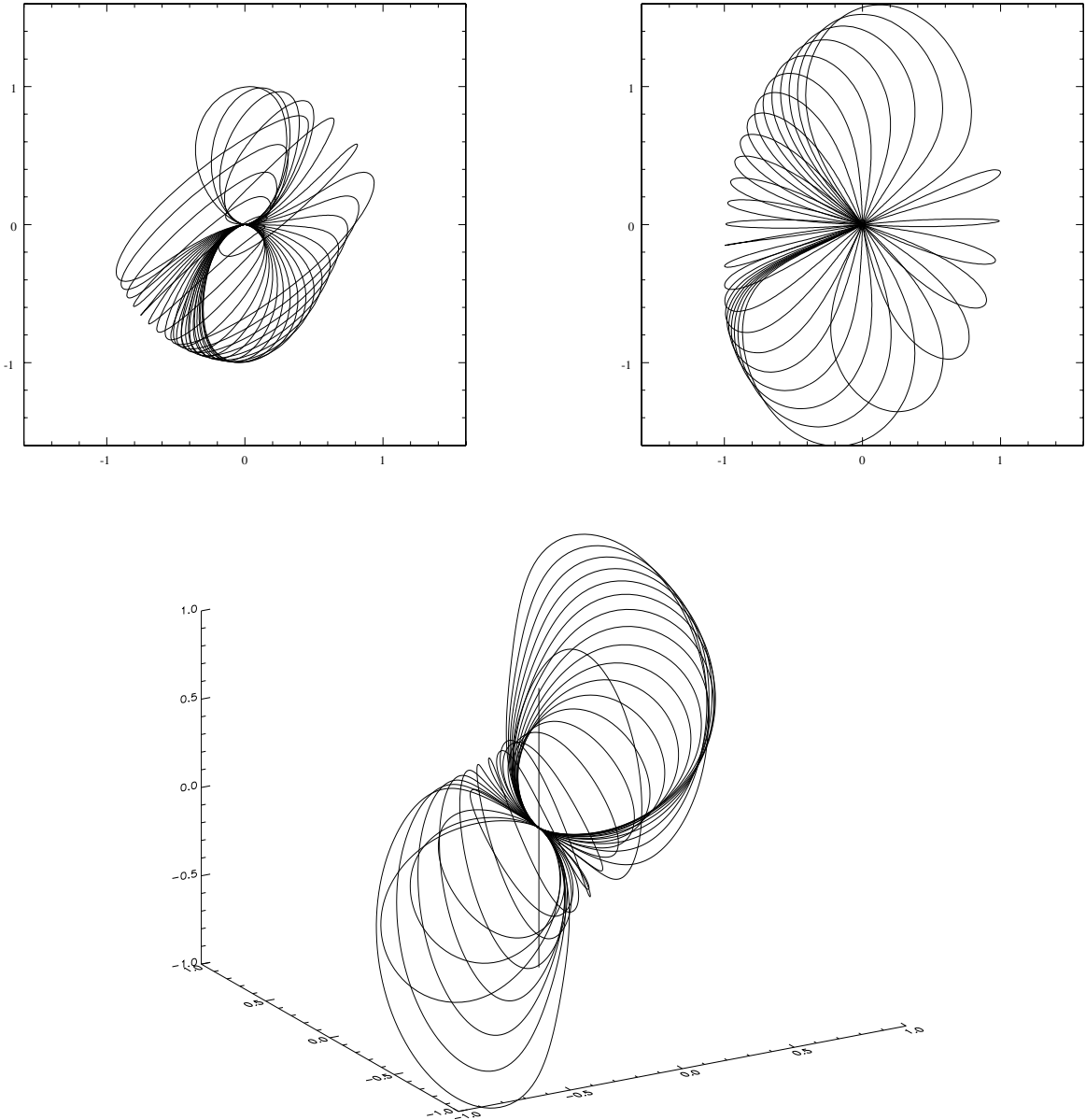,height=9cm,width=12cm}
}
\vspace{-0.12in}
\caption{The last closed field lines for a dipole inclined at
$\alpha = 60^{\circ}$ to the rotation axis.  The layout is the same as
in Figure 1.   The distortions are quite strong at this
inclination. This figure can be compared to Figure 1 of Romani \&
Yadigaroglu (1995), who show a $70^{\circ}$ inclination.}
\end{figure}

\newpage
%  90 degrees double figure
\begin{figure}
\vspace{-0.2in}
\centerline{
\psfig{figure=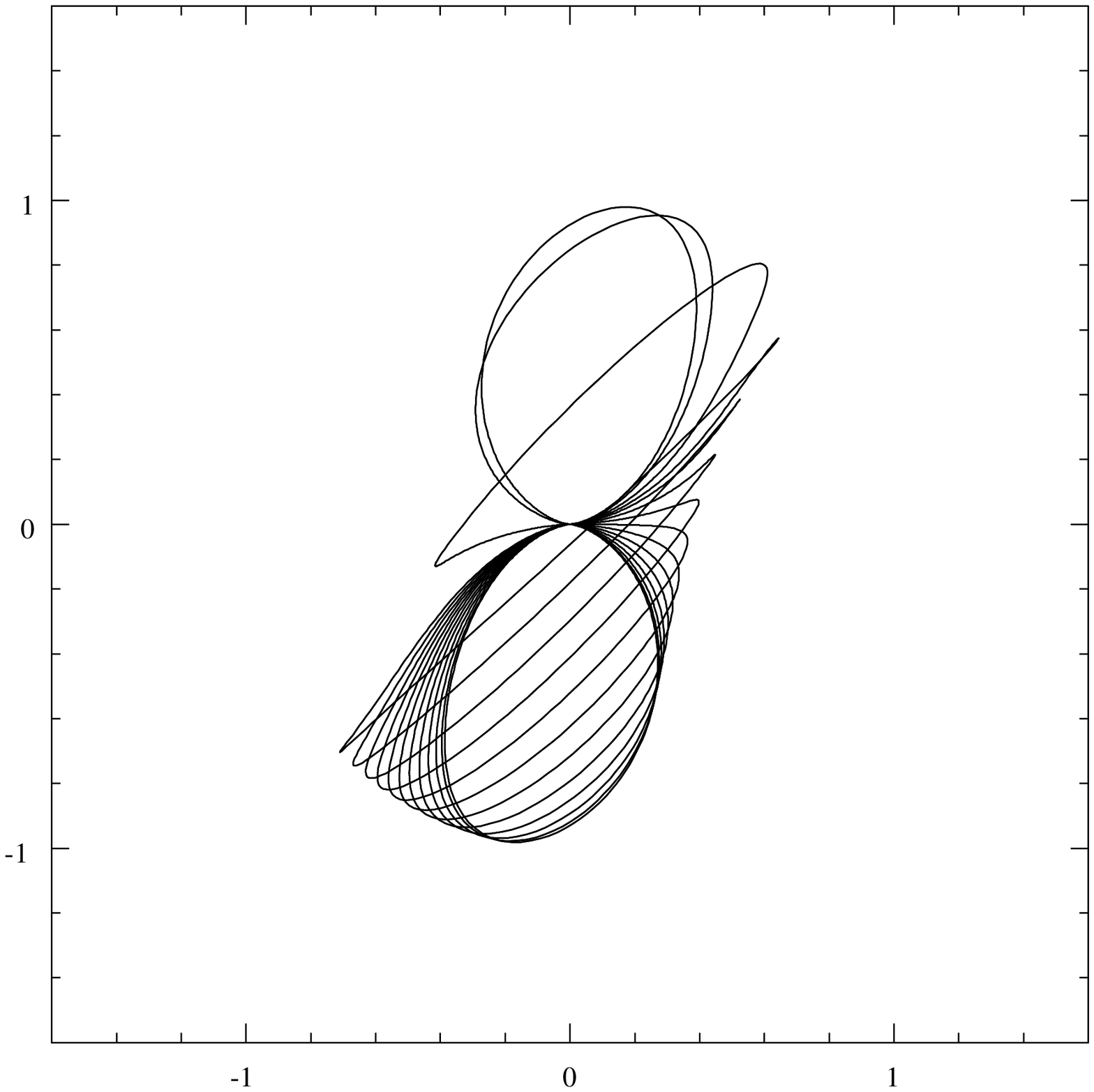,height=7.5cm,width=7.5cm}
\hspace{0.4in}
\psfig{figure=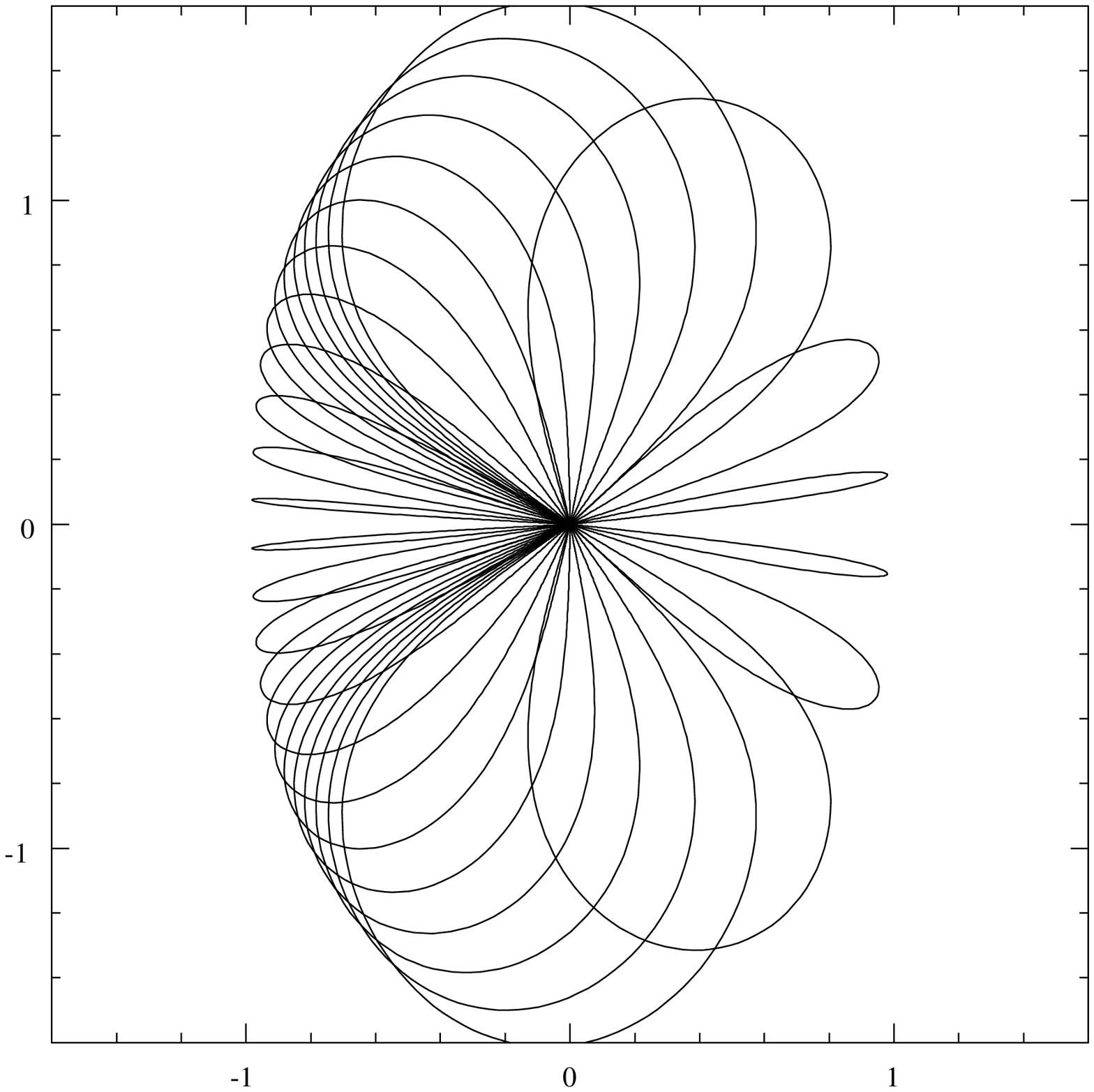,height=7.5cm,width=7.5cm}
}
\vspace{-0.12in}
\end{figure}
\begin{figure}
\vspace{-0.8in}
\centerline{
\psfig{figure=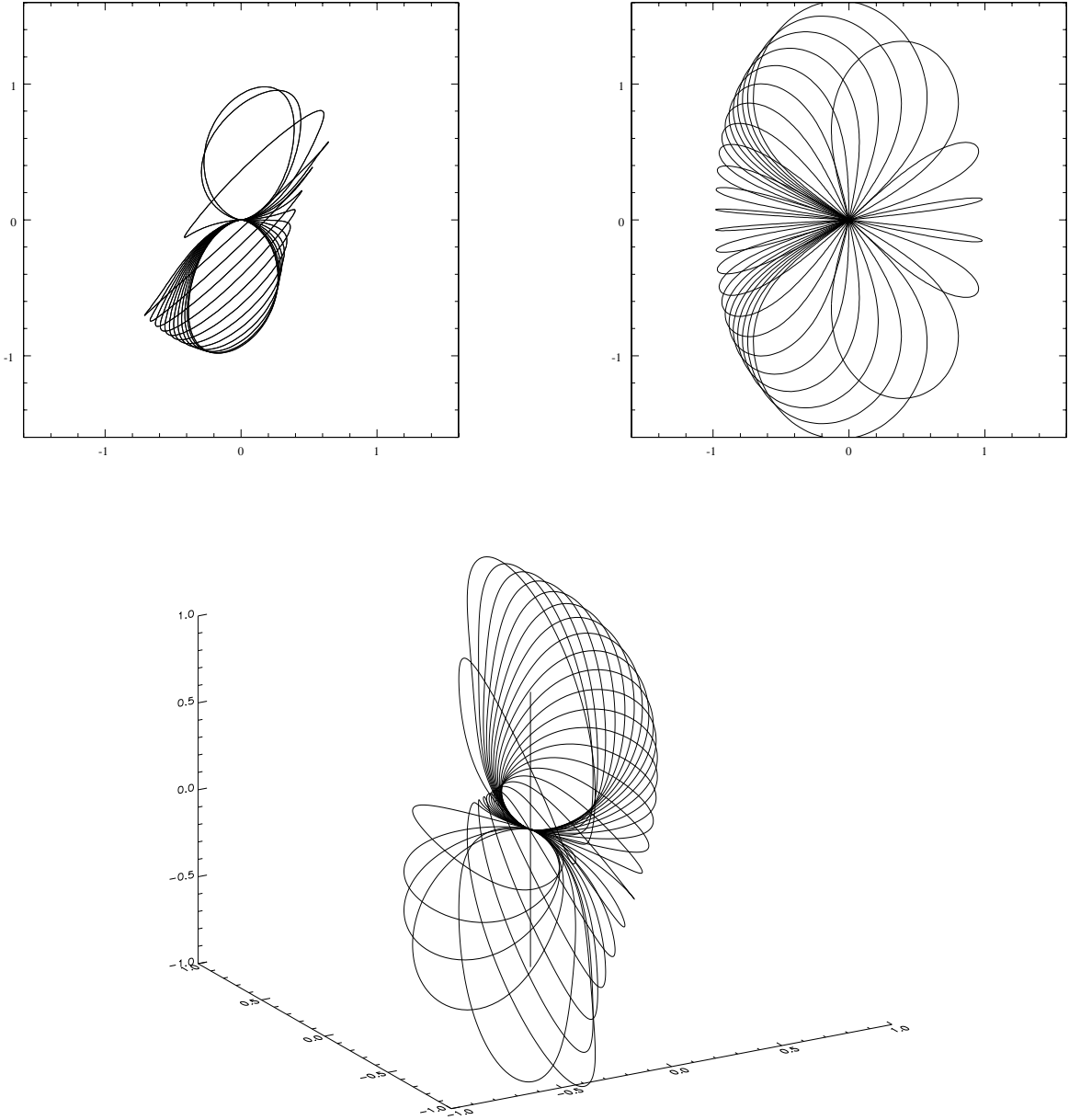,height=9cm,width=12cm}
}
\vspace{-0.12in}
\caption{The last closed field lines for a dipole inclined at
$\alpha = 90^{\circ}$ to the rotation axis.  The layout is the same as
in Figure 1.  This figure can be compared to Figure 1 of Higgins and
Henriksen (1997), who show an orthogonal rotator.}
\end{figure}

%  and consequences for polar cap outline
\begin{figure}
\vspace{-.8in}
\centerline{
\psfig{figure=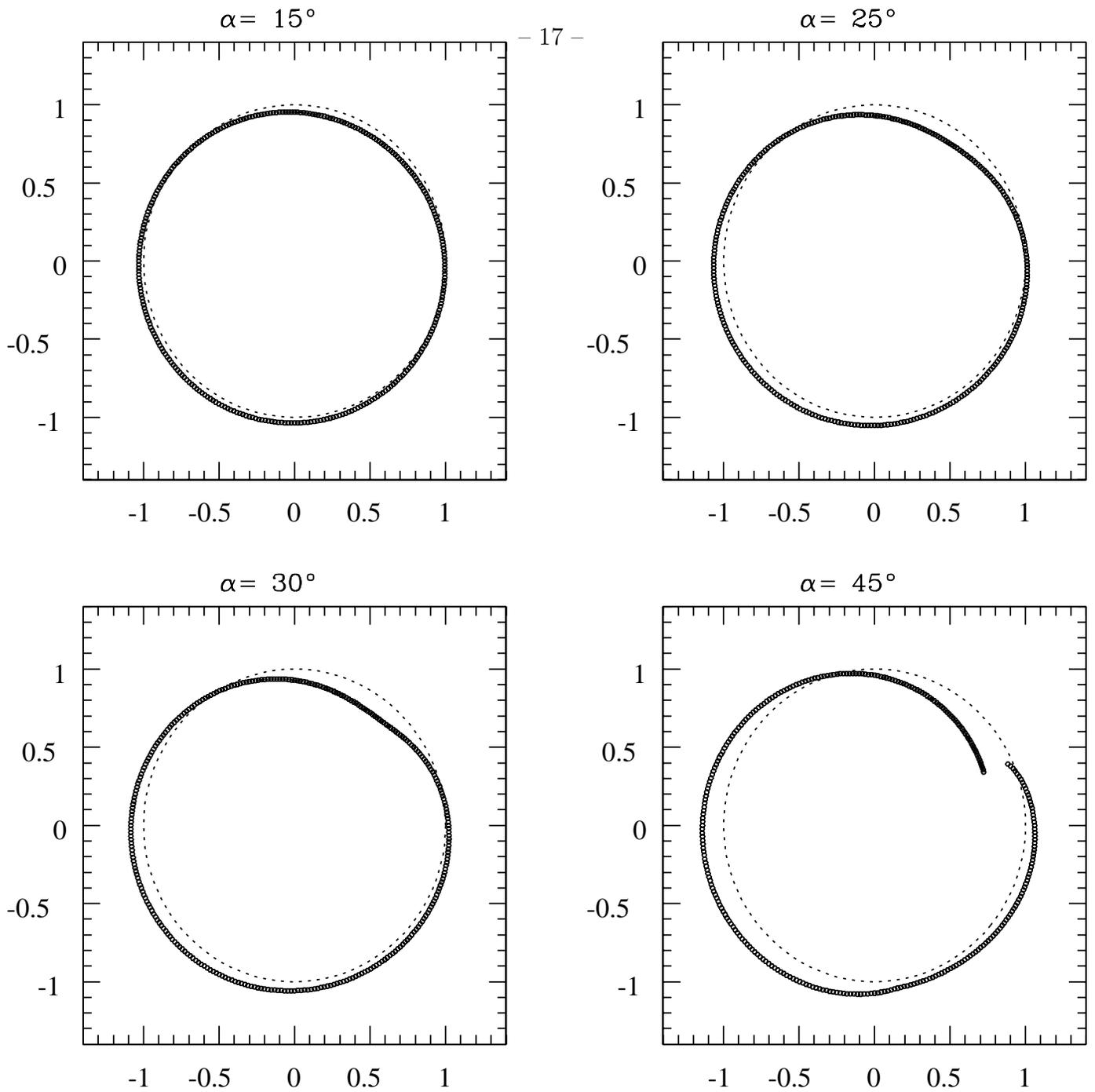}
}
\vspace{-0.12in}
\caption{The shape of the polar cap for four low inclination
angles:  $\alpha = 15^{\circ}, 25^{\circ}, 30^{\circ}$ and
$45^{\circ}$. The dashed line shows the classical, circular polar cap,
given by $\rho_{PC}$.  The heavy line shows the footpoints of field
lines which actually close at the light cylinder. Rotation is to the
right, and the magnetic axis is to the top, in these figures.}
\end{figure}

\begin{figure}
\vspace{-0.8in}
\centerline{
\psfig{figure=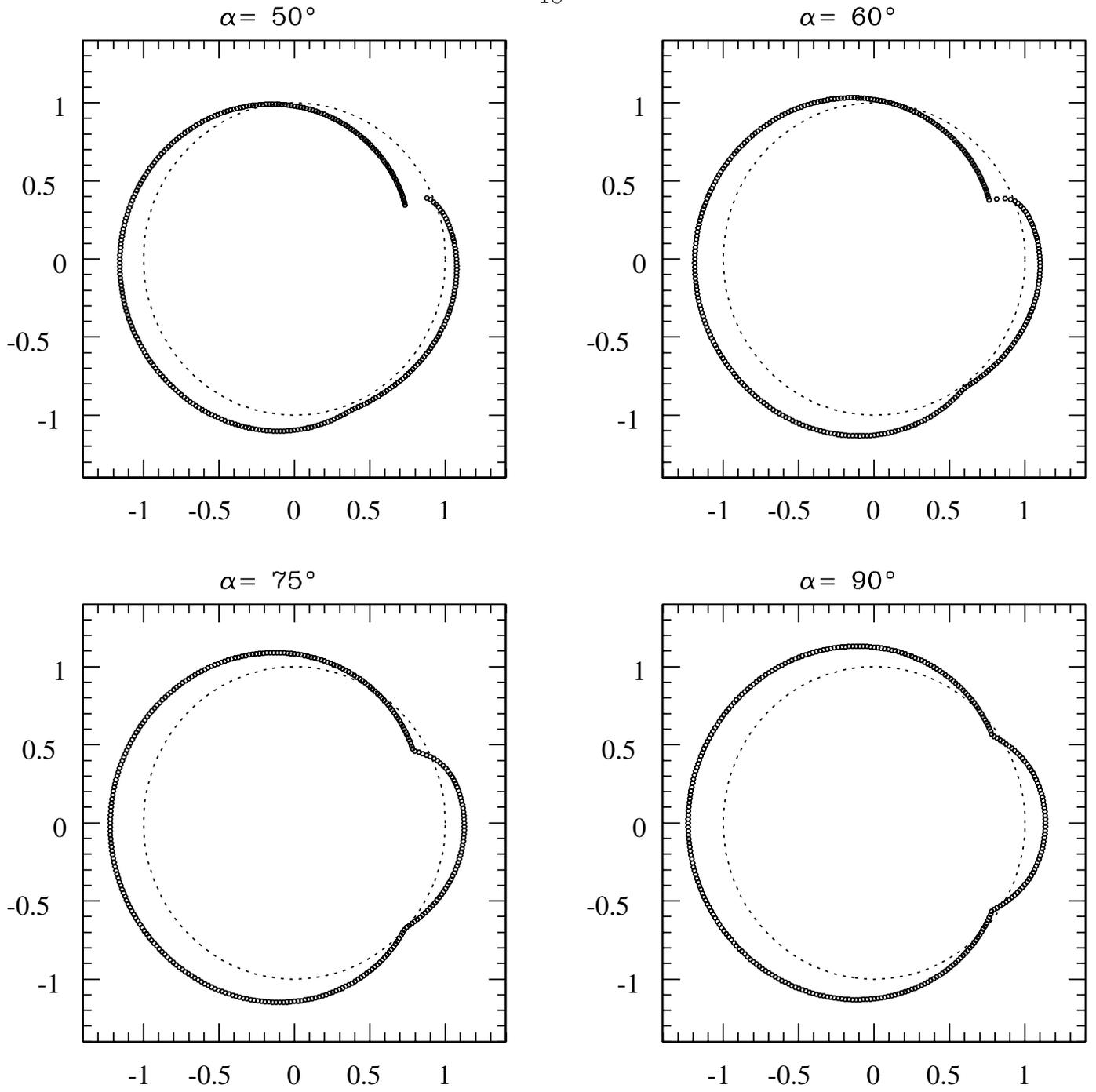}
}
\vspace{-0.12in}
\caption{The shape of the polar cap for four large inclination
angles:  $\alpha = 50^{\circ}, 60^{\circ}, 75^{\circ}$ and
$90^{\circ}$. The layout is the same as in Figure 5.}
\end{figure}

\end{document}